# Direct three-dimensional measurement of refractive index via dual photon-phonon scattering

Antonio Fiore[1], Carlo Bevilacqua[1,2,+] and Giuliano Scarcelli[1*]

[1]*Fischell Department of Bioengineering, University of Maryland, College Park, MD 20742, USA*
[2]*Dipartimento Interateneo di Fisica, Università degli Studi di Bari, I-70126 Bari, Italy*

*e-mail address: scarc@umd.edu
[+]Current affiliation: European Molecular Biology Laboratory (EMBL), Heidelberg, Germany

We developed a microscopy technique that can measure the local refractive index without sampling the optical phase delay of the electromagnetic radiation. To do this, we designed and experimentally demonstrated a setup with two co-localized Brillouin scattering interactions that couple to a common acoustic phonon axis; in this scenario, the ratio of Brillouin frequency shifts depends on the refractive index, but not on any other mechanical and/or optical properties of the sample. Integrating the spectral measurement within a confocal microscope, the refractive index is mapped at micron-scale three-dimensional resolution. As the refractive index is probed in epi-detection and without assumptions on the geometrical dimensions of the sample, this method may prove useful to characterize biological cells and tissues.

When light propagates inside a material, the phase of the electromagnetic wave is sensitive to the optical path, which fundamentally couples the geometrical path and the local index of refraction. Thus, methods to map the refractive index of a material (e.g. phase contrast microscopy [1,2], digital holography microscopy [3,4], optical coherence tomography [5,6], etc. [7–10]) are intrinsically indirect as they rely on the knowledge, assumption, or measurement of the spatial dimensions of the sample. Even when optical path delay and thickness of the sample are well characterized, we can only obtain the average refractive index along the beam propagation axis. This fundamental issue has important practical ramifications as it prevents performing spatially-resolved measurements of the index of refraction. Mapping the distribution and variations of the local refractive index is potentially crucial to analyze mass density behavior in cell biology [11–14], cancer pathogenesis [13,15,16], and corneal or lens refraction [17–19].

Several techniques in the past years have emerged to circumvent the coupling of geometrical path and refractive index in 3D samples using a tomographic approach, i.e. performing multiple measurements from different angles to reconstruct the internal refractive index distribution [20–24]. Tomographic phase microscopy enabled spatially-resolved measurements of refractive index for the first time. However, as the individual measurements are still based on optical path delay, known geometrical boundary conditions and/or reference refractive index values are needed as well as access to the sample from at least two sides. In addition, even under these conditions, the measurements are subject to artifacts due to phase wrapping when phase variations inside the sample are not smooth [1,2,13].

Here we present a novel microscopy technique that probes the refractive index of materials relying on photon-phonon interactions, not optical path delays, and thus decouples optical from geometrical path. The technology is based on measuring two inelastic Brillouin scattering interactions in confocal configuration, probing the same acoustic phonon axis so that the refractive index is the only physical quantity that affects the ratio of the two Brillouin frequency shifts. We experimentally demonstrate that using this dual photon-phonon scattering, the refractive index can be determined directly, with three-dimensional spatial resolution, accessing the sample from a single side and without assumptions on the geometrical dimensions of the sample.

To understand the light-matter interaction governing this phenomenon, we shall use the classical inelastic light scattering formalism (see Supplementary Section I for full derivation) [25,26]. Let's consider an incident electric field $\mathbf{E_i}(\mathbf{r},t) = \mathbf{E_{i0}} e^{i(\mathbf{k_i}\cdot\mathbf{r}-\omega_i t)}$ + c.c., where $\mathbf{k_i}$ is the wavevector, $\omega_i$ the frequency, $\mathbf{r}$ is the scattering position vector, and $\mathbf{E_{i0}}$ expresses field amplitude and polarization. If $\mathbf{E_i}$ encounters a fluctuation of the dielectric constant tensor $\delta\boldsymbol{\varepsilon}(\mathbf{r},t)$, the resulting scattered electric field $\mathbf{E_s}$ can be obtained from Maxwell equations in dielectric media:

$$\nabla^2 \mathbf{E_s} - \frac{1}{c^2}\frac{\partial^2 \mathbf{E_s}}{\partial t^2} = \frac{1}{\varepsilon_0 c^2}\frac{\partial^2 \mathbf{P}}{\partial t^2} \qquad (1)$$

where $\mathbf{P}(\mathbf{r},t)$ is the induced additional polarization in the medium:

$$\mathbf{P}(\mathbf{r},t) = \frac{1}{4\pi}\delta\boldsymbol{\varepsilon}(\mathbf{r},t)\cdot\mathbf{E_i}(\mathbf{r},t) \qquad (2)$$

Eq. (1) and (2) are generally valid for scattering phenomena and show how the additional polarization acts as a source term dictating the emitted scattered field. For Brillouin light scattering of phonons, the variation of the dielectric constant is induced by the acoustic displacement in the medium $\mathbf{u}(\mathbf{r},t)$ due to spontaneous thermally-driven collective sound waves inside material which obey the wave equation and are characterized by speed v and attenuation parameter Γ [26]. Here, we will focus on the longitudinal acoustic modes, since they induce significantly more efficient light scattering [27]. Longitudinal modes result in negligible depolarization so that the polarization of the scattered field is the same as the incident field [26]. In these conditions, the additional polarization can be written as:

$$\mathbf{P}(\mathbf{r},t) \propto \int |d\mathbf{q}|(\mathbf{p}\,\nabla\mathbf{u}(\mathbf{q}))\,\mathbf{E}_{i0}e^{i[(\mathbf{k}_i-\mathbf{q})\cdot\mathbf{r}-(\omega_i-\Omega(\mathbf{q}))t]} +$$
$$\int |d\mathbf{q}|(\mathbf{p}\,\nabla\mathbf{u}(\mathbf{q}))^*\,\mathbf{E}_{i0}e^{i[(\mathbf{k}_i+\mathbf{q})\cdot\mathbf{r}-(\omega_i+\Omega(\mathbf{q}))t]} + c.c. \quad (3)$$

where $\mathbf{p}$ is the elasto-optic tensor that links a certain $\delta\varepsilon(\mathbf{r},t)$ to the strain $\nabla\mathbf{u}(\mathbf{r},t)$ and we expressed the acoustic waves in terms of their spatial Fourier components, with $\mathbf{q}$ and $\Omega$ wavevector and the frequency of the acoustic wave, respectively.

The additional polarization in Eq. (3) will give rise to two scattered electric field contributions, respectively Stokes and anti-Stokes components, with oscillating frequency of $\omega_s = \omega_i \pm \Omega(\mathbf{q})$. For acoustic phonons, $\Omega \ll \omega_i$, thus we can consider the acoustic displacement a weak function of time compared to the electric field and write $\partial^2\mathbf{P}/\partial t^2 \sim \mathbf{P}$. Moreover, given the dispersion relationship $k = \omega n(\omega)/c$, we can also approximate $|\mathbf{k}_i - \mathbf{k}_s| \approx 2k_i \sin(\theta/2)$, where $\theta$ is the scattering angle, i.e. the angle between $\mathbf{k}_i$ and $\mathbf{k}_s$. In these conditions, a solution to Eq. (1) for the scattered field is:

$$\mathbf{E}_s(\mathbf{r}',t) \propto$$
$$\int |d\mathbf{q}|\,\mathbf{p}\,\nabla\mathbf{u}(\mathbf{q})\mathbf{E}_{i0}e^{i(\mathbf{k}_s\cdot\mathbf{r}'-(\omega_i-\Omega(\mathbf{q}))t)}\int |d\mathbf{r}|e^{i[(\mathbf{k}_i-\mathbf{k}_s-\mathbf{q})\cdot\mathbf{r}]}$$
$$+ \int|d\mathbf{q}|\,(\mathbf{p}\nabla\mathbf{u}(\mathbf{q}))^*\mathbf{E}_{i0}e^{i(\mathbf{k}_s\cdot\mathbf{r}'-(\omega_i+\Omega(\mathbf{q}))t)}\int|d\mathbf{r}|e^{i[(\mathbf{k}_i-\mathbf{k}_s+\mathbf{q})\cdot\mathbf{r}]}$$
$$+ c.c. \quad (4)$$

where $\mathbf{r}'$ is the observation position vector, generally of much larger amplitude than the scattering position vector ($r' \gg r$), so that the unit vector $\hat{\mathbf{k}}_s \approx (\mathbf{r}' - \mathbf{r})/r'$.

In the limit of no acoustic and optical attenuation, the quantity $(\mathbf{k}_i - \mathbf{k}_s \pm \mathbf{q})$ is real and the integration in $|d\mathbf{r}|$ yields a delta function $\delta(\mathbf{k}_i - \mathbf{k}_s \pm \mathbf{q})$. The delta function and the oscillating frequency of $\mathbf{E}_s$ are often interpreted as the conservation of momentum and energy of the phenomenon:

$$\mathbf{k}_s = \mathbf{k}_i \pm \mathbf{q} \qquad \omega_s = \omega_i \pm \Omega \quad (5)$$

Since a given observation position $\mathbf{r}'$ defines a wavevector $\mathbf{q}$ through the momentum conservation relation, we can study the electric field in the wavevector domain. In practice, every material will present both optical and acoustic attenuations, so that the quantity $(\mathbf{k}_i - \mathbf{k}_s \pm \mathbf{q})$ is complex, which leads to a frequency spread of the scattered light for any given wavevector $\mathbf{q}$. Considering the complex dispersion relationships for acoustic waves [28] and electromagnetic waves in optically isotropic material (see Suppl. Section III for anisotropic materials), the scattered electric field reads:

$$\mathbf{E}_s(\mathbf{q},t) \propto$$
$$\frac{-ie^{i(\mathbf{k}_s\cdot\mathbf{r}'-(\omega_i-\Omega)t)}\mathbf{p}\nabla\mathbf{u}(\mathbf{q})\mathbf{E}_{i0}e^{i(\mathbf{k}_i-\mathbf{k}_s-\mathbf{q})\cdot\mathbf{r}}}{\left(\frac{2n_1\omega_i}{c}\sin\left(\frac{\theta}{2}\right)-\frac{\Omega}{v(\hat{\mathbf{q}},\Omega)}\right)+i\left(\frac{2n_2\omega_i}{c}\sin\left(\frac{\theta}{2}\right)-\frac{\Gamma(\hat{\mathbf{q}},\Omega)q^2}{2v(\hat{\mathbf{q}},\Omega)}\right)}$$
$$+ AS + c.c. \quad (6)$$

where we have expressed the refractive index $n(\omega) = n_1(\omega) + in_2(\omega)$ in terms of its real, $n_1$, and imaginary, $n_2$, parts and AS stands for the anti-Stokes term.

As expected, the difference in frequency between the incident and the scattered light corresponds to the frequency of the acoustic wave. From the first term of the denominator in Eq. (6) we can obtain an expression for the frequency shift as function of incident frequency ($\omega_i$), scattering geometry ($\theta$), refractive index ($n_1$) and speed of sound (v):

$$\Omega = Re(q)\,v(\hat{\mathbf{q}},\Omega) = 2\frac{\omega_i}{c}n_1(\omega_i)v(\hat{\mathbf{q}},\Omega)\sin\frac{\theta}{2} \quad (7)$$

Finally, we can calculate the power spectrum of the scattered light using the Wiener-Kintchin theorem:

$$S(\mathbf{q},\omega) \propto \frac{\frac{\Gamma(\hat{\mathbf{q}},\Omega)q^2}{2}+\frac{n_2\Omega}{n_1}}{[\omega-(\omega_i\pm\Omega)]^2+\left[\frac{\Gamma(\hat{\mathbf{q}},\Omega)q^2}{2}+\frac{n_2\Omega}{n_1}\right]^2} \quad (8)$$

Equation (8) shows the expression of the Brillouin spectrum featuring two Lorentzian peaks, Stokes and anti-Stokes components, symmetric with respect to the incident frequency and with peaks shifted by $\pm\Omega$.

In our method, we use two scattering geometries: 1) back-scattering, with incident and scattered photons normal to the sample interface, (0-geometry, Fig. 1a); 2) symmetric angled configuration with incident and scattered photons at an angle $\alpha$ with the sample interface ($\alpha$-geometry, Fig. 1b).
Assuming negligible acoustic dispersion in the Brillouin frequency range, the frequency shifts for these geometries can be written as:

$$\Omega_0 = \omega_i\frac{2n_1(\omega_i)v(\hat{\mathbf{q}}_0)}{c} \quad (9)$$

$$\Omega_\alpha = \omega_i\frac{2n_1(\omega_i)v(\hat{\mathbf{q}}_\alpha)}{c}\left(1-\frac{\sin^2\alpha}{n_1^2(\omega_i)}\right)^{\frac{1}{2}} \quad (10)$$

As shown in Fig. 1, these configurations probe the same phonon axis, i.e. $\hat{\mathbf{q}}_0 = \hat{\mathbf{q}}_\alpha \equiv \hat{\mathbf{q}}$. Thus, defining $R = \Omega_\alpha/\Omega_0$ as the ratio between frequency shifts, we can derive the following expression for the local index of refraction:

$$n_1(\omega_i) = \left(\frac{\sin^2\alpha}{1-R^2}\right)^{\frac{1}{2}} \quad (11)$$

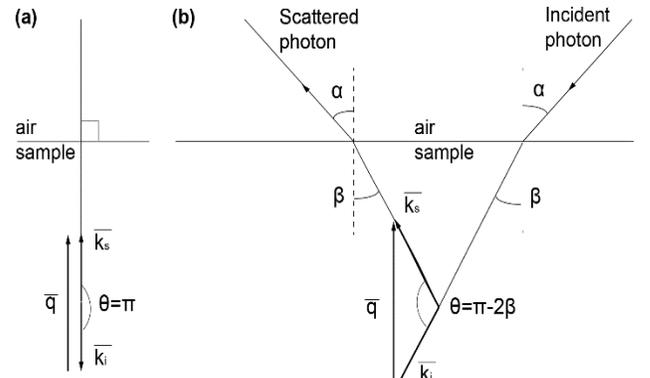

**Figure 1.** (a) 0-geometry configuration; (b) $\alpha$-geometry configuration. Both geometries sample the same acoustic phonon direction inside the volume of interaction.

In Eq. (11), the dependence on other quantities of Eq. (7) is lost because the wavelength is the same for both geometries, and the speed of the phonon is the same in both configurations, as it is sampled along a common axis. Therefore, the ratio between the two Brillouin shifts depends only on the local index of refraction of the sample inside the scattering volume of interaction. Since no other physical quantity influences the measurement, this can be considered a direct measurement of the refractive index.

To demonstrate our method, we built the experimental setup in Fig. 2a; a 532nm continuous wave laser is split in two beams of different diameter: 5.4mm and 1.8mm. The incident beams are both focused into the sample through a dry 20x 0.75NA objective lens (Supplementary Section VII). The first beam is on the optical axis, i.e. at 0° incidence angle and its corresponding scattered light is also collected on the optical axis. The second incident beam is off-axis at 5.5mm from the center of the lens at an angle α=37° and its corresponding scattered light is collected symmetrically at the same angle α. The scattered light for both geometries is coupled into the same single mode optical fiber acting as confocal pinhole, thus collecting light only from a microscopic 3D region within the sample. To facilitate collection of scattered photons from the proper scattering geometry, the two beams are alternated by a system of two shutters and a flip mirror (Fig. 2a). It would be technically straightforward to perform simultaneous measurements with two couplers and two spectrometers, or a parallelized spectral measurement [29]. A representative set of spectra for both geometries is shown in Fig. 2b.

To measure the small frequency shift due to photon-phonon interaction we built a double-stage VIPA spectrometer with ~0.7 GHz linewidth and 6 dB insertion losses (see Supplementary Section IV). The VIPA etalon is conceptually like a Fabry-Perot etalon but has a front surface with a highly reflective coating and a narrow stripe anti-reflection coated to enable light input. Tilting the etalon enables single shot spectral measurements with acquisition times in the order of 0.1-1s [30–32]. By scanning the sample with a translational stage and measuring two spectra at each location, the refractive index is mapped with 3D resolution. However, because of the different illumination/collection paths [33], the voxels sampled by the two geometries do not fully overlap. The different voxel size worsens the effective resolution of the measurement which we define as the smallest region of homogeneous refractive index required to obtain a measurement without artifacts. In our configuration, the 0-geometry has a voxel of 1.31 x 1.31 x 33 μm; the α-geometry at an angle of ~37° has a voxel of 5.3 x 3.9 x 5.7 μm. Thus, our overall resolution is 5.3 μm x 3.9 μm (lateral) x 33 μm (axial). Improvements in spatial resolution can be achieved with different beam diameter and objective lens configurations (see Supplementary Section V).

We validated our index of refraction measurement against gold-standard Abbe refractometry. We prepared several solutions of water and sodium chloride (NaCl) from pure water to the saturation point. All solutions were prepared and tested at constant temperature of 22 °C. The different salt concentration induces a variation of the refractive index from 1.33 to 1.39. Our method is highly consistent with the gold standard as the correlation is linear, the slope is ~1 with $R^2$>0.99 (Fig. 3).

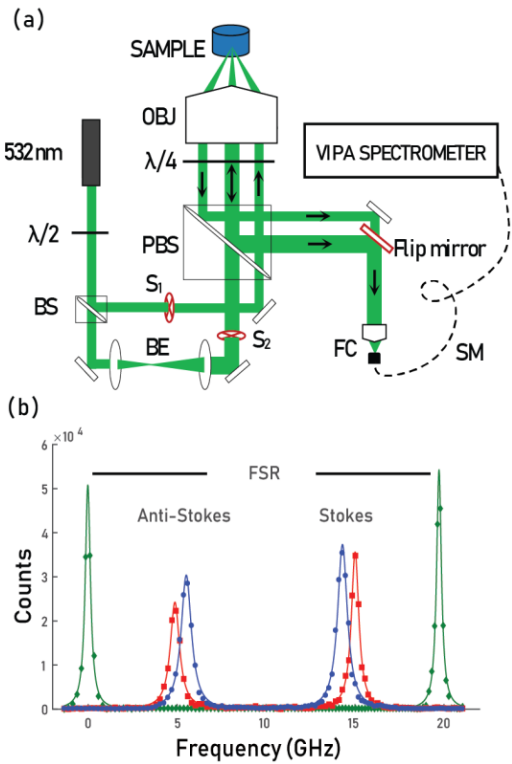

**Figure 2.** (a) Schematic of the dual-geometry Brillouin spectroscopy. A flip mirror and two shutters (S1, S2) work in alternate configuration to allow acquisition of the two scattering geometries. (b) Methanol spectrum for 0-geometry (blue) and α-geometry (red); the Rayleigh scattering peak (green) shows the free spectral range (FSR) of the etalon. Experimental data (dots) are fitted to a double Lorentzian curve (solid line) in MATLAB.

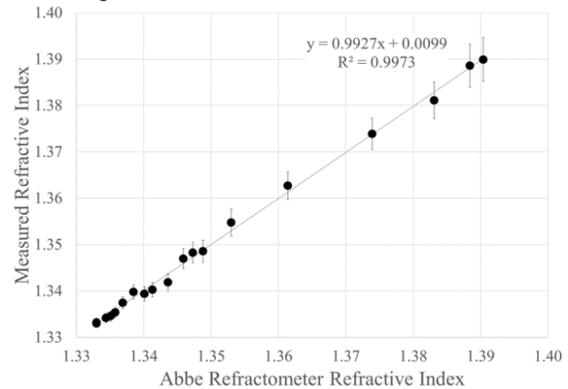

**Figure 3.** Measured refractive index for different water-NaCl solutions compared with Abbe refractometer's values. The larger error at high salt concentration is due to Brillouin peak broadening that affects the accuracy of the fit.

To demonstrate our 3D mapping capability, we fabricated a non-symmetric sample using a drop of cured optical glue (Thorlabs, NOA61, n=1.56), deposited on a glass-bottom petri dish and surrounded by methanol (n=1.329). The result of our refractive index mapping of XY and XZ sections are in Figure 4. For this image, we used 15 mW incident power, and acquisition time 1 and 3s for backscattering and α-angle geometry, respectively. Despite the high asymmetry along the z-axis, we obtained accurate values for the index of refraction and we reliably reconstructed the dome profile of the deposited drop. A sample like this would be difficult to characterize for techniques based on optical path delay, in particular without access to the sample from two sides.

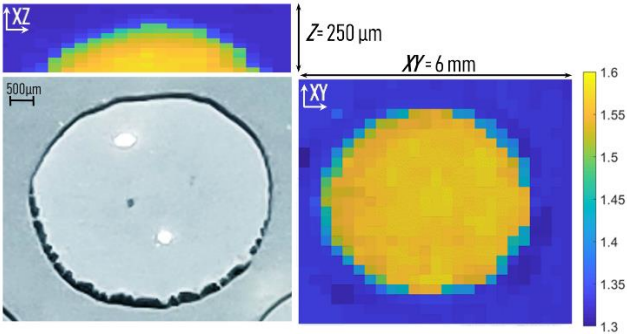

**Figure 4.** XY and XZ cross sections of a drop of photo-activated polymer (n=1.56) surrounded by methanol (n=1.329). Bright field image is also reported for visual comparison.

Interestingly, the co-localized Brillouin interaction can provide information also about the speed of the acoustic wave and the imaginary part of the refractive index. Indeed, from Eq. (8), the linewidth of the Lorentzian peaks is proportional to the attenuation coefficients of electro-magnetic and acoustic waves. In a material with no optical absorption, the linewidth is given by $\Gamma q^2$, i.e. the inverse of the phonon lifetime; with strong optical absorption, the electric field amplitude decreases exponentially so that the imaginary part of the refractive index is dominant [34]. In the general case, both terms contribute to the linewidth of the spectrum; with our dual scattering geometry, the two terms can be independently measured if the spectrometer has sufficient spectral resolution (Supplementary Section II).

In summary, we have demonstrated that the dual photon-phonon scattering measurement of the frequency shift can provides the real part of the index of refraction $n_1$ and the speed of sound v, while the measurement of the linewidth in the two geometries provides the imaginary part of the refractive index $n_2$ and the sound attenuation coefficient $\Gamma$. These quantities can be measured directly and locally with micrometric three-dimensional resolution within a standard confocal microscope in epi-detection configuration

The possibility of retrieving material refractive index using photon-phonon scattering had been proposed three decades ago [35–38]. Our approach brings three crucial innovations to earlier methods. First, at a fundamental level, we devised the co-localization of two scattering geometries to probe the same phonon axis so that anisotropies in the physical properties of the sample do not affect the refractive index measurement. Second, our confocal sampling of the probed volume of interaction provides mapping capabilities at high 3D resolution. Third, our common lens configuration only needs to access samples from one side. From a practical standpoint, we have also developed much faster Brillouin spectrometer that enable mapping of the refractive index.

The uncertainty on the refractive index measurement can be determined from Eq. (11). One potential source of uncertainty is the evaluation of the angle α, but the angle can be calibrated with high precision. Here, we fitted α value using three reference materials of known refractive index and Brillouin shift; other methods to determine the angle of incidence of the α-geometry beam can be implemented [39]. The uncertainty on the refractive index can be expressed as linear function of the shift measurement precision (Supplementary Section VI). In our experiment, we reached a refractive index precision of ~0.001 using a laser power of 15mW and exposure time of ~s. We operated in shot noise conditions, thus the frequency shift precision improves with the square root of the signal-to-noise ratio.

A potential source of error comes from the asymmetric broadening of the Brillouin linewidth due to the spread of angle illuminated and collected by the objective lens. In our low NA conditions, such broadening would induce an estimation error of ~0.3% of the shift but was avoided by experimental calibration with materials of known Brillouin shift. At higher NA, this error may increase, however it can be effectively eliminated either by modeling the broadening term for known illuminating-collecting geometry [40], or by adjusting the Lorentzian fit to dismiss asymmetric broadening artifacts [41]. Another potential source of error comes from the acoustic dispersion of the material. The derivation of Eq. (11) assumes constant speed of sound at the frequencies of the two scattering geometries and thus it needs to be modified if acoustic dispersion is significant. However, a linear behavior, i.e. no acoustic dispersion, in the gigahertz frequency range has been reported in many liquid and solid materials [42–49].

In conclusion, we have reported a novel method that allows direct mapping of refractive index in confocal configuration. Our instrument can map the index of refraction inside a non-symmetric structure with 3D micron-level resolution.

The authors thank G. Di Caprio and E. Edrei for helpful discussions. This is work was supported by NIH (R33CA204582, R01EY028666, R01HD095520, U01CA202177), NSF (CMMI-1537027) and University of Bari (Global Thesis Scholarship to CB).